# Optimization of thermal comfort in building through envelope design

*Milorad Bojič[a], Alexandre Patou-Parvedy[b], Harry Boyer[c]*

[a] *Faculty of Engineering, Kragujevac, Serbia, milorad.bojic@gmail.com*
[b] *Faculty of Engineering, Kragujevac, Serbia, parvedyalexandre@gmail.com,*
[c] *LPBS, Equipe Physique et ingénierie mathématique appliquée à l'énergie et l'environnement,* University of Reunion Island, Reunion, France, *Harry.Boyer@univ-reunion.fr*

**Abstract:**
Due to the current environmental situation, energy saving has become the leading drive in modern research. Although the residential houses in tropical climate do not use air conditioning to maintain thermal comfort in order to avoid use of electricity. As the thermal comfort is maintained by adequate envelope composition and natural ventilation, this paper shows that it is possible to determine the thickness of envelope layers for which the best thermal comfort is obtained. The building is modeled in EnergyPlus software and HookeJeves optimization methodology. The investigated house is a typical residential house one-storey high with five thermal zones located at Reunion Island, France. Three optimizations are performed such as the optimization of the thickness of the concrete block layer, of the wood layer, and that of the thermal insulation layer. The results show optimal thickness of thermal envelope layers that yield the maximum TC according to Fanger predicted mean vote.

# 1. Introduction

Due to the current environmental situation, energy saving has become the leading drive in modern research. To limit electricity consumption in tropical regions, it is important to achieve the optimal thermal comfort in residential houses without using electricity. It is important in tropical Reunion as in the last fifteen years, the consumption of energy has been multiplied by 2. Reunion demography has significantly increasing and will continue to increase, and the energy demand will depend on it.

Research in the optimization of thermal comfort was performed by Kurian et al [1], Magnier and Haghighat[2], Chantrelle et al [3], and Stavrakakis et al[4]. Kurian et al. showed that dimming with a fuzzy logic based window blind controller provides complete optimization of thermal comfort in the interior with energy savings. To optimize thermal comfort, Magnier and Haghighat performed multiobjective optimization of building design HVAC system settings, thermostat programming, and passive solar design. Chantrelle et al developed of a multicriteria tool for optimizing the renovation of buildings where one of criteria was the optimized thermal comfort. Stavrakakis et al optimized window-openings design for thermal comfort in naturally ventilated buildings. However, the literature does not report the research on optimization of thermal comfort by the adequate selection of width of layers in the building envelope construction.

In this research, the objective is to enhance the thermal comfort in a house located in Reunion Island without using electricity for air conditioning. In this direction, an optimization of the thickness of house walls has been done. The optimized layers of the wall are the customary materials used in Reunion Island namely concrete block, thermal insulation (glass wool) and wood. Three optimizations are performed such as the optimization of the thickness of the concrete block layer, of the wood layer, and that of the thermal insulation layer. In these optimizations, other two layers are kept at constant width. These optimizations yield the thickness of each layer of the wall for the best thermal comfort inside the house. The optimization is performed by programming at GenOpt code where thermal comfort of the entire house as an objective function is maximized. This code is coupled with an external simulation program (EnergyPlus).

## 2. Description

### 2.1 House

In this study, the residential house has a floor surface of 162 m². The house consists of 2 bedrooms and, a kitchen, a living room and a toilet. A family of 4 peoples: 2 children and 2 parents live in the house. For each room, a occupancy schedule is recorded. The parents work from 8 am to 6.30 pm. They spend their lunch break at the house from 12 am to 1.30 pm. During the week, the children are at school from 8 am to 5.30 pm. For the weekend, the entire family stays at the house. The infiltration is assumed to be 0.5 ach throughout these optimizations.

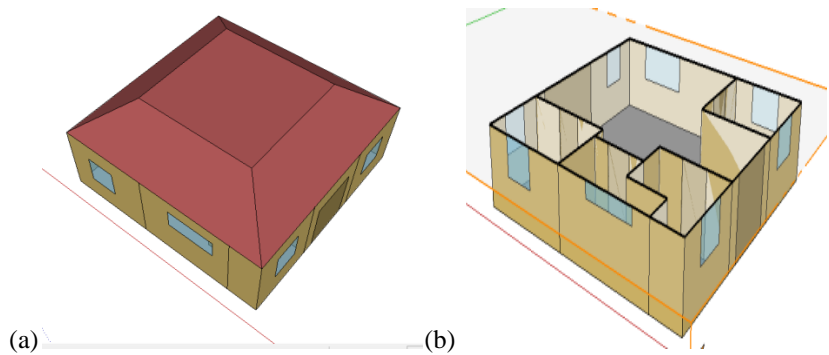

(a)　　　　　　　　　　(b)

Fig. 1 The investigated Reunion residential house: (a) entire house, b) cut through the house

The wall layers in this house are of customary materials used in the wall construction in Reunion Island. From the outside to the inside, the first layer is of wood, the second of glass wool, and the third of concrete block. Their thickness, thermal conductivity, density and specific heat are given in Table 1. During optimization, the thickness of each layer of the wall is varied in an interval from 0.001m to 1m.

The concrete blocks are covered by a mortar layer of 1 mm and the wood by a varnish layer of 0.1 mm. These two layers are not considered in these optimizations.

Reunion is a hot and humid tropical island that lies to 21° south latitude and 55° east longitude. Reunion Island has a yearly mean temperature of between 26ºC to 27ºC and has high daytime temperatures of 29°C to 34°C and relative humidity of 70% to 90% throughout the year.

Table 1 Characteristics of the layers of the wall of the investigated house

|   |               | Thickness | k      | $\rho$ | c       |
|---|---------------|-----------|--------|--------|---------|
|   |               | m         | W/m-K  | kg/m3  | J/kg-m3 |
| 1 | Wood          | 0.025     | 0.15   | 608    | 1630    |
| 2 | Glass wool    | 0.025     | 0.04   | 11     | 800     |
| 3 | Concrete block| 0.2032    | 1.11   | 800    | 920     |

**2.2 Software**

The optimization is made with GenOpt software. GenOpt is an optimization program for the minimization of a cost function that is evaluated by an external simulation program, such as EnergyPlus, TRNSYS, SPARK, IDA-ICE or DOE-2. It has been developed for optimization problems where the cost function is computationally expensive and its derivatives are not available or may not even exist. GenOpt can be coupled to any simulation program that reads its input from text files and writes its output to text files. The independent variables can be continuous variables (possibly with lower and upper bounds), discrete variables, or both. Constraints on dependent variables can be implemented using penalty or barrier functions.

**2.3Thermal comfort**

The thermal comfort climate is defined as "that state of mind which expresses satisfaction with the thermal environment" [5]. So, thermal comfort is very difficult to objectively quantify because it relies on a wide range of environmental and personal factors that decides on what will make people feel thermally comfortable.

Thermal comfort is maintained when the heat generated by human metabolism is allowed to dissipate, thus maintaining thermal equilibrium with the surroundings. It has been long recognized that the sensation of feeling hot or cold is not just dependent on air temperature alone.

At present, two different approaches for the definition of thermal comfort coexist, each one with its potentialities and limits: the rational or heat-balance approach and the adaptive approach. The rational approach uses data from climate chamber studies to support its theory, best characterized by the works of Fanger while the adaptive approach uses data from field studies of people in building. Fanger's Predicted Mean Vote (PMV) model was developed in the 1970's from laboratory and climate chamber studies. In these studies, participants were dressed in standardised clothing and completed standardised activities, while exposed to different thermal environments. In some studies the researchers chose the thermal conditions, and participants recorded how hot or cold they felt, using the seven-point ASHRAE thermal sensation scale shown in Table 2. In other studies, participants controlled the thermal environment themselves, adjusting the temperature until they felt thermally 'neutral' (i.e. neither hot nor cold; equivalent to voting '0' on the ASHRAE thermal sensation scale).

Table.2: Seven point thermal sensation scale [6]

| Sensation | Description |
|---|---|
| 3 | Hot |
| 2 | Warm |
| 1 | slightly warm |
| 0 | neutral |
| -1 | slightly cool |
| -2 | cool |
| -3 | cold |

Maintaining this heat balance is the first condition for achieving a neutral thermal sensation. However, Fanger noted that ''man's thermoregulatory system is quite effective and will therefore create heat balance within wide limits of the environmental variables, even if comfort does not exist''.

To be able to predict conditions where thermal neutrality would occur, Fanger investigated the body's physiological processes when it is close to neutral.

That comfort equation was expanded using data from 1296 participants. The resulting equation described thermal comfort as the imbalance between the actual heat flow from the body in a given thermal environment and the heat flow required for optimum (i.e. neutral) comfort for a given activity. This expanded equation related thermal conditions to the seven-point ASHRAE thermal sensation scale, and became known as the ''Predicted Mean Vote'' (PMV) index.

The PMV index suggested by Fanger predicts the mean response of a large group of people according to the ASHRAE thermal sensation scale.

## 2.4 Objective function

The objective function that was minimised is the total *PMV* given as

$$PMV_{total} = C_{main\ room}\ PMV_{main\ room} + C_{kitchen}\ PMV_{kitchen} + C_{bedroom1}\ PMV_{bedroom1} + C_{bedroom2}\ PMV_{bedroom2} \quad (8)$$

where $C_{main\ room} = 0.4169$ is the weight coefficient for the part of the time that the family spends in the main room during one year, $C_{kitchen} = 0.3533$ is the weight coefficient for the part of the time that the family spends in the kitchen during one year, $C_{bedroom1} = 0.1896$ is the weight coefficient for the part of the time that the family spends in the bedroom 1 during one year, $C_{bedroom1} = 0.0402$ is the weight coefficient for the part of the time that the family spends in the bedroom 2 during one year. The part of the time that family spends in a room is a ratio of the time that the family spends in one room, and the total time the family spends in the entire house. The PMV for each zone is the output of EnergyPlus for the modelled house.

## 3. Results and discussion
The purpose of this research is to improve (decrease) the PMV (the thermal comfort) in the house for the optimized house. Three optimization runs are performed to find the compositions of the wall that give the best thermal comfort. As the external wall has three layers, Three optimizations are performed such as the optimization of the thickness of the concrete block layer, of the wood layer, and that of the thermal insulation layer. Each optimization run gives the optimum width of different layer while keeping other two at the constant width.

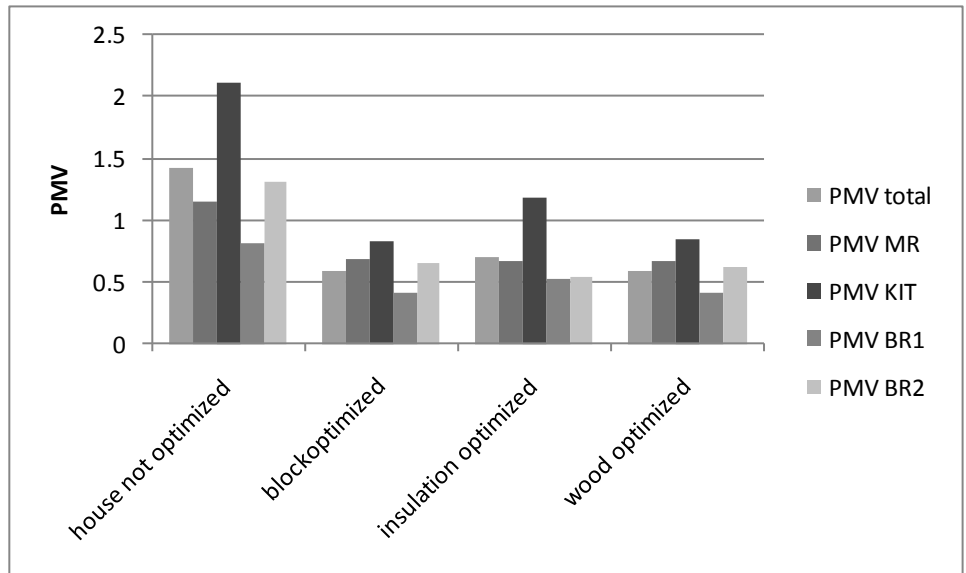

Fig.3: Comparison of PMV between house no optimized and house with different layer optimized

Figure 3 shows the total PMV and the PMV for each room in the house without optimization and three houses with optimization of a wall layer. Generally, the house with the optimized wall layer has a better PMV than the house without any optimization. The PMV of the house without any optimization is around 1.45. With the optimization, this value drops to about 0.5, by 34.5 %.

The house with the optimized block layer seems to be the best constructive solution. The total PMV is around 0.55 instead to of 1.45 for house without an optimization. Then, the PMV of each room drops significantly to be around 0.5. In the main room (MR), the PMV is around 0.7 instead 1.15 for the house without optimization. This drop by 60.86 % gives a neutral sensation of thermal comfort for the inhabitant instead previous slight warm sensation. The most noticeable drop in PMV is in the kitchen (KIT). In the house without optimization, the PMV is around 2.10 which gives a warm sensation of thermal comfort. After the optimization the PMV drops to 0.82.

The house with the optimized wood layer is also a good constructive solution. In comparison with the previous case, the PMV in the kitchen is a little bit higher and is around 0.85. For the rest of rooms, the PMV is around 0.5 that is similar to the previous case.

The house with the optimized insulation layer is the worst solution. Even if the total PMV and that for the others rooms yield to a neutral sensation of comfort, the PMV of the kitchen is 1.17 meaning that occupants in kitchen have slightly warm sensation of thermal comfort.

Figure 4 shows the thickness of the wall layers for a house without and with three optimizations. The house without optimization has the thickness of concrete blocks of 160 mm, of thermal insulation (glass wool) of 25mm, and wood of 25mm. However the thickness of the wall varies for each optimization run. For the house where the concrete block is optimized, the thickness of the concrete block is 200 mm which is 40mm thicker than that for the house without optimization. The thickness of the entire wall is now 250 mm instead 210mm-the rise of 19%. For the house where the wood is optimized, the wood layer thickness is 100 mm instead of 25mm for the house without optimization. The wood layer is normally used as a decoration so the optimized layer width seems to be very thick for the practical applications. In the house where the thermal insulation layer is optimized, the thermal insulation width is also 100 mm instead 25mm. The total thickness of the wall in each case goes up. In these two cases of the optimization, the total thickness of the wall is 285 mm that is a rise of 36%.

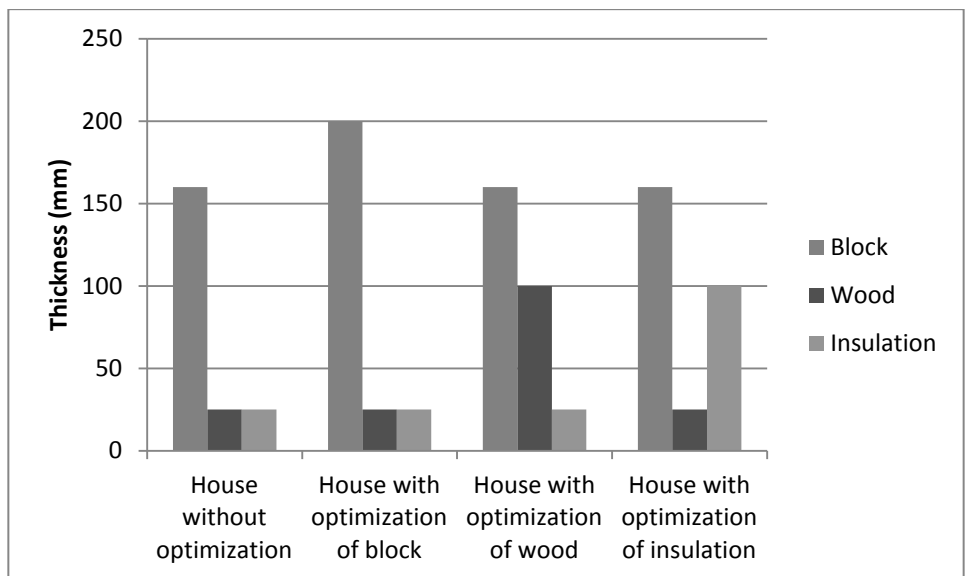

Fig.4: Thickness of the wall layers for a house with optimizations and without optimization

## 3. Conclusion

The optimization of the thickness of the wall gives better thermal comfort during the time that the family is in the house. The total PMV drops by 34.5% and the PMV of each room by about 60%.

The thermal comfort is better when the wall is thicker. In fact in the initial house, the thickness of the wall was 210 mm and in the optimized house the wall is equal and over 250 mm.

The best case is when the concrete block is optimized. Then, the thickness of the wall rises by 40mm (19%), the total PMV is around 0.55 and the PMV of each room gives a neutral sensation of comfort. The optimized thermal insulation gives a good PMV, however its thickness is increased by 300%. The optimized wood layer give the worst results.

# Acknowledgment

This paper is a result of two project investigations: (1) project TR33015 of Technological Development of Republic of Serbia, and (2) project III 42006 of Integral and Interdisciplinary investigations of Republic of Serbia. The first project is titled "Investigation and development of Serbian zero-net energy house", and the second project is titled "Investigation and development of energy and ecological highly effective systems of poly-generation based on renewable energy sources. We would like to thank to the Ministry of Education and Science of Republic of Serbia for their financial support during these investigations.